\documentclass[11pt]{article}
\usepackage{hyperref}
\usepackage{algorithm,algorithmic}
\usepackage{amsfonts,amsmath, amssymb, bm}

\def\frac#1#2{{#1\over #2}}

\def\x{{\mathbf x}}
\def\y{{\mathbf y}}

\def\dotfil{\leaders\hbox to 1.5mm{.}\hfill}
\newcounter{rmnum}
\def\RN#1{\setcounter{rmnum}{#1}\uppercase\expandafter{\romannumeral\value{rmnum}}}
\def\rn#1{\setcounter{rmnum}{#1}\expandafter{\romannumeral\value{rmnum}}}

\newcommand{\FNorm }[1]{\mbox{}\left\|#1\right\|_F  }
\newcommand{\FNormS}[1]{\mbox{}\left\|#1\right\|_F^2}

\newcommand{\TNorm }[1]{\mbox{}\left\|#1\right\|_2  }
\newcommand{\TNormS}[1]{\mbox{}\left\|#1\right\|_2^2}

\newcommand{\setlinespacing}[1]%
           {\setlength{\baselineskip}{#1 \defbaselineskip}}

\newcommand{\abs }[1]{\left|#1\right|}

\newtheorem{lemma}{Lemma}
\newtheorem{theorem}{Theorem}
\newtheorem{corollary}{Corollary}

\newenvironment{Proof}{\noindent {\em Proof:}}{\\\hspace*{\fill}\mbox{$\diamond$}}

\newcommand{\mat}[1]{{\ensuremath{\bm{\mathrm{#1}}}}}

\def\exp{\hbox{\rm exp}}

\def\e{{\mathbf e}}

\def\x{{\mathbf x}}

\def\matA{\mat{A}}
\def\matB{\mat{B}}

\def\matI{\mat{I}}

\def\matM{\mat{M}}

\def\matX{\mat{X}}

\begin{document}
\title{A Note on Randomized Element-wise Matrix Sparsification}

\author{
Abhisek Kundu
\thanks{
Department of Computer Science,
Rensselaer Polytechnic Institute,
Troy, NY,
kundua2@rpi.edu.
}
\and
Petros Drineas
\thanks{
Department of Computer Science,
Rensselaer Polytechnic Institute,
Troy, NY, drinep@rpi.edu.
}
}

\date{}
\maketitle

\begin{abstract}
\noindent Given a matrix $\matA \in \mathbb{R}^{m \times n}$, we present a randomized algorithm that sparsifies $\matA$ by retaining some of its elements by sampling them according to a distribution that depends on both the square and the absolute value of the entries. We  combine the ideas of \cite{DZ11, AKL13} and provide an elementary proof of the approximation accuracy of our algorithm following \cite{DZ11} without the truncation step.
\end{abstract}
\section{Introduction}
Element-wise matrix sparsification was pioneered in~\cite{AM01,AM07} and was later improved in~\cite{DZ11,AKL13}. More specifically, the original work of~\cite{AM01,AM07} sampled entries from a matrix with probabilities depending on the square of an entry for ``large'' entries and on the absolute value of an entry for ``small'' entries. \cite{DZ11} proposed to zero out the small entries and then used sampling with respect to the squares of the remaining entries in order to sparsify the matrix; an elegant proof was possible via a matrix-Bernstein inequality. Very recently,~\cite{AKL13} argued that the zeroing out step could be avoided by sampling with respect to the absolute values of the matrix entries. Theorem~\ref{lem:elementsampling} combines the ideas of~\cite{DZ11,AKL13} to provide an elementary proof that bypasses the zeroing out step. More specifically, we avoid zeroing out the small elements of the input matrix by constructing a sampling probability distribution that depends on both the absolute values \textit{as well as} the squares of the entries of the input matrix.

\section{Our Result}

We present our main algorithm (Algorithm~1) and the related Theorem~\ref{lem:elementsampling}, which is our main quality-of-approximation result for Algorithm~1.

\subsection{Notation}

We use bold capital letters (e.g., $\matX$) to denote matrices and bold lowercase letters (e.g., $\x$) to denote column vectors. Let $[n]$ denote the set $\{1, 2, ..., n\}$. We use $\mathbb{E}(X)$ to denote the expectation of a random variable $X$; when $\matX$ is a random matrix, $\mathbb{E}(\matX)$ denotes the element-wise expectation of each entry of $\matX$. For a matrix $\matX \in \mathbb{R}^{m \times n}$, the Frobenius norm $\FNorm{\matX}$ is defined as $\FNormS{\matX} = \sum_{i,j=1}^{m,n} \matX_{ij}^2$, and the spectral norm $\TNorm{\matX}$ is defined as $\TNorm{\matX} = \text{max}_{\TNorm{\y}=1}\TNorm{\matX\y}$. For symmetric matrices $\matA, \matB$ we say that $\matB \succeq \matA$ if and only if $\matB - \matA$ is a positive semi-definite matrix. $\matI_n$ denotes the $n \times n$ identity matrix and $\ln x$ denotes the natural logarithm of $x$. Finally, we use $\e_i$ to denote standard basis vectors whose dimensionalities will be clear from the context.

\subsection{Algorithm}

Our main algorithm (Algorithm 1) randomly samples (in independent, identically distributed trials) $s$ elements of a given matrix $\matX$ according to a probability distribution $\{p_{ij}\}_{i,j=1}^{m,n}$ over the elements of $\matX$.

\begin{algorithm}\label{alg:alg1}
\centerline{\caption{Element-wise Matrix Sparsification Algorithm}}
\begin{algorithmic}[1]
\STATE \textbf{Input:} $\matX \in \mathbb{R}^{m \times n}$, $\left\{p_{ij}\right\}_{i,j=1}^{m,n}$ such that $p_{ij} \geq 0$ (for all $i,j$) and $\sum_{i,j=1}^{m,n} p_{ij}=1$, integer $s > 0$.
\STATE \textbf{For} $t = 1\ldots s$ (i.i.d. trials with replacement) \textbf{randomly sample} pairs of indices $(i_t, j_t) \in \{1\ldots m\}\times \{1\ldots n\}$ with
$\mathbb{P}\left[ (i_t, j_t) =  (i,j)\right]\ =\ p_{ij}.$
\STATE \textbf{Output:} set of sampled pairs of indices $\Omega = \left\{\left(i_t,j_t\right),t=1\ldots s\right\}.$
\STATE \textbf{Sampling operator:} $\mathcal{S}_{\Omega}: \mathbb{R}^{m \times n}\rightarrow \mathbb{R}^{m \times n}$ with
$\mathcal{S}_{\Omega}\left(\matX\right) = \frac{1}{s}\sum_{t=1}^s \frac{\matX_{i_t j_t}}{p_{i_t j_t}} \e_{i_t} \e_{j_t}^T.$
\end{algorithmic}
\end{algorithm} 

\begin{theorem}\label{lem:elementsampling}
Let $\matX \in \mathbb{R}^{m \times n}$ and  let $\epsilon > 0$ be an accuracy parameter. Let $\mathcal{S}_{\Omega}: \mathbb{R}^{m \times n} \rightarrow \mathbb{R}^{m \times n}$ be the sampling operator of the element-wise sampling algorithm (Algorithm 1) and assume that the sampling probabilities $\left\{p_{ij}\right\}_{i,j=1}^{m,n}$ satisfy
\begin{equation}\label{eqn:probability}
p_{ij} \geq \frac{\beta}{2}\left(\frac{\matX_{ij}^2}{\FNormS{\matX}} + \frac{\abs{\matX_{ij}}}{\sum_{i,j=1}^{m,n}\abs{\matX_{ij}}}\right)
\end{equation}
for all $i,j$ and some $\beta \in (0,1]$. Then, with probability at least $1 - \delta$,
$$\TNorm{\mathcal{S}_{\Omega}(\matX) - \matX} \leq \epsilon,$$
%
if either \textbf{(i)} $\epsilon \leq \FNorm{\matX}$ \textbf{and} $s \geq  \frac{6 \max\{m,n\} \ln\left(\left(m+n\right)/\delta\right)}{\beta \epsilon^2} \FNormS{\matX}$,

or \textbf{(ii)} $\epsilon > \FNorm{\matX}$ \textbf{and} $s \geq  \frac{6 \max\{m,n\} \ln\left(\left(m+n\right)/\delta\right)}{\beta \epsilon} \FNorm{\matX}$.\\
\end{theorem}

\noindent We now restate the above bound in terms of the stable rank of the input matrix. Recall that the stable rank is defined as $\mbox{\textbf{sr}}\left(\matX\right):=\FNormS{\matX}/\TNormS{\matX}$ and is upper bounded by the rank of $\matX$.
\begin{corollary}
Let $\matX \in \mathbb{R}^{m \times n}$, let $\epsilon > 0$ be an accuracy parameter such that $\mbox{\textbf{sr}}\left(\matX\right) \geq \epsilon^2$, and let $\mathcal{S}_{\Omega}(\matX)$ be the sparse sketch of $\matX$ constructed via Algorithm~1 with the $p_{ij}$'s satisfying the bounds of eqn.~(\ref{eqn:probability}). If
$$s \geq \frac{6 \max\{m,n\}\ln\left(\left(m+n\right)/\delta\right)}{\beta \epsilon^2} \mbox{\textbf{sr}}\left(\matX\right),$$
then, with probability at least $1 - \delta$,
$$\TNorm{\matX - \mathcal{S}_{\Omega}(\matX)} \leq \epsilon\TNorm{\matX}.$$
\end{corollary}
\section{Proof of Theorem~\ref{lem:elementsampling}}\label{sxn:appendix:primitive2}

In this section we provide a proof of Theorem~\ref{lem:elementsampling} following the lines of~\cite{DZ11}. First, we rephrase the non-commutative matrix-valued Bernstein bound theorem of~\cite{Recht09} using our notation.
\begin{theorem}\label{theorem:Recht}
[Theorem 3.2 of \cite{Recht09}] Let $\matM_1, \matM_2, ..., \matM_s$ be independent, zero-mean random matrices in $\mathbb{R}^{m \times n}$. Suppose $\max_{t\in [s]} \left\{\TNorm{\mathbb{E}(\matM_t\matM_t^T)}, \TNorm{\mathbb{E}(\matM_t^T\matM_t)} \right\} \leq \rho^2$ and $\TNorm{\matM_t} \leq \gamma$ for all $t \in [s]$. Then, for any $\epsilon > 0$,
$$\TNorm{\frac{1}{s}\sum_{t=1}^{s}\matM_t} \leq \epsilon$$
holds, subject to a failure probability of at most
$$\left ( m+n\right )\exp\left ( \frac{-s\epsilon^2/2}{\rho^2 + \gamma\epsilon /3} \right ).$$
\end{theorem}
For all $t \in [s]$ we define the matrix $\matM_t \in \mathbb{R}^{m \times n}$ as follows:
\begin{eqnarray}\label{eqn:Mt}
\matM_t = \frac{\matX_{i_tj_t}}{p_{i_tj_t}}\e_{i_t}\e_{j_t}^T - \matX.
\end{eqnarray}
It now follows that
$$\frac{1}{s}\sum_{t=1}^{s}\matM_t = \frac{1}{s}\sum_{t=1}^{s}\left[ \frac{\matX_{i_tj_t}}{p_{i_tj_t}}\e_{i_t}\e_{j_t}^T - \matX \right ] = S_{\Omega}(\matX) - \matX.$$
Let $\textbf{0}_{m \times n}$ denote the $m \times n$ all-zeros matrix and note that $\matX = \sum_{i,j=1}^{m,n}\matX_{ij}\e_{i}\e_{j}^T$. The following derivation is immediate (for all $t \in [s]$):
$$\mathbb{E}(\matM_t) = \mathbb{E}\left(S_{\Omega}(\matX)\right) - \matX = \sum_{i,j=1}^{m,n}p_{ij}\frac{\matX_{ij}}{p_{ij}}\e_{i}\e_{j}^T - \matX = \textbf{0}_{m \times n}.$$
The next lemma bounds $\TNorm{\matM_t}$ for all $t \in [s]$.
\begin{lemma}\label{lemma:boundMt}
Using our notation, $\TNorm{\matM_t} \leq \frac{3\sqrt{mn}}{\beta} \FNorm{\matX}$ for all $t \in [s]$.
\end{lemma}
\begin{Proof}
Notice that sampling according to the element-wise probabilities of eqn.~(\ref{eqn:probability}) satisfies
$$p_{ij} \geq \frac{\beta}{2}\frac{\abs{\matX_{ij}}}{\sum_{i,j=1}^{m,n}\abs{\matX_{ij}}}.$$
We can use the above inequality to get
\begin{eqnarray*}
\TNorm{\matM_t} =  \TNorm{\frac{\matX_{i_tj_t}}{p_{i_tj_t}}\e_{i_t}\e_{j_t}^T - \matX}
\leq \frac{2}{\beta}\sum_{i=1}^{m}\sum_{j=1}^{n}\abs{\matX_{ij}} +\TNorm{\matX} \leq \frac{3\sqrt{mn}}{\beta} \FNorm{\matX}.
\end{eqnarray*}
In the above we used $\beta\leq 1$, $\TNorm{\matX} \leq \FNorm{\matX}$, and (from the Cauchy-Schwarz inequality) $$\sum_{i,j=1}^{m,n}\abs{X_{ij}} \leq \sqrt{mn \sum_{i,j=1}^{m,n} X_{ij}^2} = \sqrt{mn} \FNorm{\matX}.$$ Thus, we get a new bound for Lemma~2 of~\cite{DZ11}, bypassing the need for a truncation step.
\end{Proof}

\noindent Next we bound the spectral norm of the expectation of $\matM_t\matM_t^T$. The spectral norm of the expectation of $\matM_t^T\matM_t$ can be bounded using a similar analysis.
\begin{lemma}\label{lemma:boundvar} Using our notation, $\TNorm{\mathbb{E}(\matM_t\matM_t^T)} \leq \frac{2n}{\beta} \FNormS{\matX}$ for all $t \in [s]$.
\end{lemma}
\begin{Proof}
Recall that $\matX = \sum_{i,j=1}^{m,n}\matX_{ij}\e_{i}\e_{j}^T$ and $\sum_{i,j=1}^{m,n}p_{ij} = 1$ to derive
\begin{eqnarray*}
\mathbb{E}[\matM_{t}\matM_{t}^T] & = & \mathbb{E}\left[ \left(\frac{\matX_{i_tj_t}}{p_{i_tj_t}}\e_{i_t}\e_{j_t}^T - \matX\right) \left(\frac{\matX_{i_tj_t}}{p_{i_tj_t}}\e_{j_t}\e_{i_t}^T - \matX^T\right) \right ]\\
& = & \sum_{i,j=1}^{m,n} p_{ij}\left(\frac{\matX_{ij}}{p_{ij}}\e_{i}\e_{j}^T - \matX\right) \left(\frac{\matX_{ij}}{p_{ij}}\e_{j}\e_{i}^T - \matX^T\right)\\
& = & \sum_{i,j=1}^{m,n}\left(\frac{\matX_{ij}^2}{p_{ij}} \e_{i}\e_{j}^T\e_{j}\e_{i}^T\right) - \left(\sum_{i,j=1}^{m,n}\matX_{ij}\e_{i}\e_{j}^T\right)\matX^T - \matX \left(\sum_{i,j=1}^{m,n}\matX_{ij}\e_{j}\e_{i}^T\right) + \sum_{i,j=1}^{m,n}p_{ij} \matX \matX^T\\
& = & \sum_{i,j=1}^{m,n}\left(\frac{\matX_{ij}^2}{p_{ij}} \e_{i}\e_{i}^T\right) - \matX \matX^T.
\end{eqnarray*}
\noindent Notice that sampling according to the element-wise sampling probabilities of eqn.~(\ref{eqn:probability}) satisfies $p_{ij} \geq \frac{\beta}{2}\frac{\matX_{ij}^2}{\FNormS{\matX}}$ and so we get
$$\mathbb{E}[\matM_{t}\matM_{t}^T] = \sum_{i,j=1}^{m,n}\left(\frac{\matX_{ij}^2}{p_{ij}}\e_{i}\e_{i}^T\right) - \matX \matX^T \preceq \frac{2\FNormS{\matX}}{\beta} \sum_{i,j=1}^{m,n}\e_{i}\e_{i}^T - \matX \matX^T = \frac{2n\FNormS{\matX}}{\beta} \matI_{m} - \matX \matX^T.$$
Using Weyl's inequality we get
\begin{eqnarray*}
\TNorm{\mathbb{E}[\matM_{t}\matM_{t}^T]} \leq \max\left\{\TNormS{\matX \matX^T}, \frac{2n\FNormS{\matX}}{\beta} \TNorm{\matI_{m}} \right\} = \frac{2n}{\beta} \FNormS{\matX}.
\end{eqnarray*}
\end{Proof}
\\
We can now apply Theorem~\ref{theorem:Recht} with $\rho^2 =  \frac{2n}{\beta} \FNormS{\matX}$ and $\gamma =  \frac{3\sqrt{mn}}{\beta} \FNorm{\matX}$ to conclude that
$\TNorm{\mathcal{S}_{\Omega}(\matX) -  \matX} \leq \epsilon$ holds subject to a failure probability at most
$$\left(m+n\right)\exp\left(\frac{-s\beta\epsilon^2}{4n\FNormS{\matX} + 2\epsilon\sqrt{mn}\FNorm{\matX}}\right).$$
Setting the failure probability equal to $\delta$, we conclude that it suffices to set $s$ as follows:
\begin{eqnarray*}
&& s \geq \frac{1}{\beta \epsilon^2}({4n\FNormS{\matX} + 2\epsilon\sqrt{mn}\FNorm{\matX}})\ln\left(\frac{m+n}{\delta}\right).
\end{eqnarray*}
We now consider two cases. First, if $\epsilon \leq \FNorm{\matX}$,
\begin{eqnarray*}
4n\FNormS{\matX} + 2\epsilon\sqrt{mn}\FNorm{\matX} &\leq& \max\{m, n\} (4\FNormS{\matX} + 2\epsilon\FNorm{\matX})\\
&\leq& 6 \max\{m, n\} \FNormS{\matX},
\end{eqnarray*}
which immediately proves the first case of Theorem~\ref{lem:elementsampling}. Similarly, if $\epsilon > \FNorm{\matX}$, 
\begin{eqnarray*}
4n\FNormS{\matX} + 2\epsilon\sqrt{mn}\FNorm{\matX} &\leq& 6 \epsilon \max\{m, n\} \FNorm{\matX}
\end{eqnarray*}
and the second case of Theorem~\ref{lem:elementsampling} follows.
\newpage \bibliography{sparsification} \bibliographystyle{plain}
\end{document}